\documentclass[prc,twocolumn,nofootinbib,superscriptaddress,showpacs]{revtex4}
\usepackage{graphicx,amsmath,amssymb,bm,multirow}
\usepackage{color}
\usepackage{amscd}

\newcommand{\be}[1]{\begin{equation}\label{#1}}
\newcommand{\ee}{\end{equation}}

\renewcommand{\vec}[1]{{\mathbf{#1}}}

\begin{document}

\title{Spurious finite-size instabilities
       in nuclear energy density functionals}

\author{V. Hellemans}
\email{vhellema@ulb.ac.be}
\affiliation{PNTPM, CP229, Universit{\'e} Libre de Bruxelles,
B-1050 Bruxelles, Belgium}

\author{A. Pastore}
\email{apastore@ulb.ac.be}
\affiliation{IAA, CP226, Universit{\'e} Libre de Bruxelles,
B-1050 Bruxelles, Belgium}
\affiliation{Universit\'e Lyon 1,
43 Bd. du 11 Novembre 1918, F-69622 Villeurbanne
cedex, France;\\ CNRS-IN2P3, UMR 5822, Institut de Physique Nucl\'{e}aire de Lyon, 4 Rue Enrico Fermi, 69622 Villeurbanne Cedex, France}

\author{T. Duguet}
\affiliation{CEA-Saclay DSM/IRFU/SPhN, F-91191 Gif sur Yvette Cedex, France}
\affiliation{National Superconducting Cyclotron Laboratory and Department of Physics and Astronomy,
Michigan State University, East Lansing, MI 48824, USA}

\author{K. Bennaceur}
\affiliation{Universit\'e Lyon 1,
43 Bd. du 11 Novembre 1918, F-69622 Villeurbanne
cedex, France;\\ CNRS-IN2P3, UMR 5822, Institut de Physique Nucl\'{e}aire de Lyon, 4 Rue Enrico Fermi, 69622 Villeurbanne Cedex, France}

\author{D. Davesne}
\affiliation{Universit\'e Lyon 1,
43 Bd. du 11 Novembre 1918, F-69622 Villeurbanne
cedex, France;\\ CNRS-IN2P3, UMR 5822, Institut de Physique Nucl\'{e}aire de Lyon, 4 Rue Enrico Fermi, 69622 Villeurbanne Cedex, France}

\author{J. Meyer}
\affiliation{Universit\'e Lyon 1,
43 Bd. du 11 Novembre 1918, F-69622 Villeurbanne
cedex, France;\\ CNRS-IN2P3, UMR 5822, Institut de Physique Nucl\'{e}aire de Lyon, 4 Rue Enrico Fermi, 69622 Villeurbanne Cedex, France}

\author{M. Bender}
\affiliation{Universit{\'e} Bordeaux,
             Centre d'Etudes Nucl{\'e}aires de Bordeaux Gradignan, UMR5797,
             F-33170 Gradignan, France}
\affiliation{CNRS/IN2P3,
             Centre d'Etudes Nucl{\'e}aires de Bordeaux Gradignan, UMR5797,
             F-33170 Gradignan, France}

\author{P.-H. Heenen}
\affiliation{PNTPM, CP229, Universit{\'e} Libre de Bruxelles,
B-1050 Bruxelles, Belgium}

\date{\today}

\begin{abstract}
\begin{description}

\item[Background]
It is known that some well-established parametrizations of the nuclear energy density
functional (EDF) do not always provide converged results for
nuclei. It has been pointed out that there is a qualitative link between
this finding and the appearance of finite-size instabilities of symmetric
nuclear matter (SNM) near saturation density when computed within the random
phase approximation (RPA).
\item[Purpose]
We seek for a quantitative and systematic connection between the
impossibility to converge self-consistent calculations of nuclei and
the occurrence of finite-size instabilities in SNM, using the example
of scalar-isovector (\mbox{$S=0$}, \mbox{$T=1$}) instabilities of the
standard Skyrme EDF. Our aim is to establish a stability criterion based
on computationally-friendly RPA calculations of SNM that is independent
on the functional form of the EDF and that can be utilized during the
adjustment of its coupling constants.
\item[Results]
Tuning the coupling constant $C^{\rho \Delta\rho}_{1}$ of the gradient term
that triggers scalar-isovector instabilities of the standard Skyrme EDF,
we find that the occurrence of instabilities in finite nuclei depends
strongly on the numerical scheme used to solve the self-consistent
mean-field equations.
The link to instabilities of SNM is made by extracting the lowest density
$\rho_{\text{crit}}$ at which a pole appears at zero energy in the RPA
response function when employing the critical value of the coupling constant
$C^{\rho \Delta\rho}_{1}$ extracted in nuclei.
\item[Conclusions]
Instabilities of finite nuclei can be artificially hidden when employing
inappropriate numerical schemes and relying on overly restrictive,
e.g.\ spherical, symmetries. Our analysis suggests a two-fold
stability criterion to avoid scalar-isovector instabilities:
(i) The density $\rho_{\text{min}}$ corresponding to the lowest pole in the RPA response function
 should be larger than about 1.2~times the
saturation density; (ii) one needs to verify that $\rho_{p}(q_{\text{pq}})$ exhibits a distinct global minimum and
 is not a decreasing function for large transferred momenta.
\end{description}

\end{abstract}


\maketitle


\section{\label{sec:intro}Introduction}

Energy density functionals (EDFs) are popular tools for the description of fermionic systems within self-consistent mean-field methods.
They are successfully applied to atomic
nuclei~\cite{bender03b}, liquid helium~\cite{barranco06a,stringari87b},
helium droplets~\cite{stringari87a,weisgerber92a}
and cold atoms in traps~\cite{Bulgac:2008tm}.

The EDF is the key ingredient of such methods. Its parameters are determined phenomenologically by fitting a few properties of selected finite as well as properties of idealized
infinite systems, such as the electron gas
or homogeneous nuclear matter.
There is, however, no certainty that an EDF constructed in this way
will not exhibit an unphysical behavior when applied to any physical system.
The example under study in this paper manifests itself by the impossibility to converge
the iterative procedure to solve the mean-field equations for the ground-states of some nuclei (e.g. see Refs.~\cite{tondeur84,lesinski06a,schunck10b,Hellemans:2011aa}).
Such non-convergence is a fingerprint of an instability of nuclear matter that can be characterized
in the idealized system of homogeneous isospin-symmetric nuclear matter (SNM).
Two cases must be distinguished.

Landau-Migdal parameters~\cite{migdal67a} can be a guide to identify instabilities in the infinite wavelength
limit (or zero momentum transfer \mbox{$\vec{q}_{\text{ph}} = \vec{0}$}) of the response
function that characterize the transition of SNM between two different homogeneous phases of matter \cite{navarro13}. State-of-the-art \textit{ab-initio} many-body calculations predict SNM to
be stable in all spin-isospin channels up to several times saturation
density~\cite{Vidana:2002pc,Bombaci:2005vi}. Therefore, Landau-Migdal parameters are often used to validate or even constrain the parameters of the EDF.

More general {\it finite-size} instabilities may appear at finite momentum transfer \mbox{$\vec{q}_{\text{ph}} \neq \vec{0}$}~\cite{lesinski06a} that cannot be monitored by Landau parameters. At a given density, those instabilities can be detected by scanning the linear response of SNM as a function of the wavelength (or transferred momentum $\vec{q}_{\text{ph}}$) of a perturbation in each of
the four spin-isospin ($S$,$T$) channels of the particle-hole interaction.
Tools for such studies have been set-up for
non-relativistic Skyrme~\cite{garciarecio92a,margueron06a,lesinski06a,Davesne:2009ky,Pastore:2012mf} parametrizations of the nuclear EDF.
The instability arises as a collective mode at zero excitation
energy, signaling that homogeneous matter undergoes a
transition to an inhomogeneous phase. The only known \textit{physical} example of such a finite-size instability
of homogeneous matter is the so-called spinodal instability in the \mbox{$S=0$},
\mbox{$T=0$} channel \cite{Pastore:2012mf,ducoin08a,ducoin08b,ducoin07},
which sets in below about 2/3 of nuclear saturation density in SNM.
It signals that homogeneous nuclear matter becomes unstable against the formation of finite-size clusters. To the best of our knowledge,
no \textit{ab-initio} prediction exists regarding the stability of SNM with
respect to finite wavelength perturbations in the three other spin-isospin channels.
However, as we shall see below, the very existence of  nuclei excludes
the possibility of finite-size instabilities of homogeneous matter
for a wide range of densities and transferred momenta.

Systematic investigations of finite-size instabilities have started only
recently~\cite{lesinski06a,Kortelainen:2010wb,schunck10b,Hellemans:2011aa,sulaksano11a}.
In the following, we limit
our\-selves to the instability related to unphysical strong oscillations of the
scalar-isovector density \mbox{$\rho_1 (\vec{r}) \equiv \rho_n (\vec{r}) - \rho_p (\vec{r})$}
that measures the local difference between neutron and proton matter density
distributions~\cite{tondeur84,lesinski06a}. Within the standard Skyrme parametrization of the nuclear EDF, such a pathology is induced by a strongly attractive
scalar-isovector gradient term of the form
\begin{equation}
E^{\rho\Delta\rho}_1 \equiv \!\int \! d^{3}\vec{r} \, {\cal E}^{\rho\Delta\rho}_1(\vec{r}) \equiv \!\int \! d^{3}\vec{r} \, C^{\rho\Delta\rho}_1 \, \rho_{1}  (\vec{r}) \, \Delta \rho_{1} (\vec{r}) \, . \label{isovectorenergy}
\end{equation}
Instabilities triggered by
terms containing gradients of spin densities have been encountered as well \cite{schunck10b,Hellemans:2011aa, pototzky10a}.

The objective of the present article is to search for a transparent, that is, quantitative and systematic, relation between the impossibility to converge self-consistent calculations of nuclei and the occurrence of finite-size instabilities in SNM, independent of the details of the EDF parametrization. Such a relationship is not straightforward because nuclei are affected by shell and surface effects that do not exist in SNM. If such a one-to-one correspondence can be established, we wish to introduce a criterion based on computationally-friendly random-phase approximation (RPA) calculations of SNM to control the stability of EDF parametrizations during the adjustment of their coupling constants.  We focus here on the standard form of the EDF \cite{bender03b} and, as mentioned above, on scalar-isovector (\mbox{$S=0$}, \mbox{$T=1$}) instabilities.

The paper is organized as follows. In Sect.~\ref{protocol}, we introduce the protocol of our analysis and detail the methods used.
The results obtained with this protocol are presented and analyzed in Sect.~\ref{results}.  Conclusions are given in Sect.~\ref{conclusions}. Two appendices provide further technical details.

\section{Protocol of the analysis}
\label{protocol}

\subsection{Procedure}

The present analysis is carried out for nine parametrizations that are all based on the traditional form~\cite{bender03b} of the Skyrme EDF. This set includes four parametrizations, KDE0v1, LNS, NRAPRii\footnote{The notation ii indicates that we have doubled the strength $W_{0}$ of the spin-orbit interaction of the original NRAPR parametrization, as suggested in \cite{stevenson12x}.}, and SQMC700
 that were shown to be consistent with a large set of pseudo-data in SNM~\cite{Dutra:2012mb} and were adjusted to no or very limited nuclear masses. In addition, we incorporated five parametrizations, SkM*, SLy5, T11, T44 and UNEDF0 in the set for which the fitting procedure includes properties of finite nuclei and which are widely used. Basic SNM properties are summarized in Table \ref{parametrizations} for the nine parametrizations.

In the standard form of the Skyrme EDF, the instability under study is governed by the term $E^{\rho\Delta\rho}_1$  and can be controlled by tuning the single parameter $C^{\rho\Delta\rho}_1$. Although the
nine EDF parametrizations  differ  in the protocol used to adjust their coupling constants, we will see that scalar-isovector instabilities are characterized by a
rather universal critical value of $C^{\rho\Delta\rho}_1$. However, as we anticipate the future use of more elaborate EDF parametrizations~\cite{Carlsson:2008gm,Stoitsov:2010ha,sadoudi11thesis} for which the appearance of an instability in a
given $(S,T)$ channel may be due to several terms in the energy density, we aim at linking the instability to a more {\it physical} quantity rather than to a
coupling constant.

\begin{table}
\begin{center}
\begin{tabular}{lcrcccc}
\hline
Param. & Ref. & $m^{*}_{s}/m$ & $m^{*}_{v}/m$ &$\rho_{sat}$ & $K_{\infty}$ &
$a_{sym} $\\
\hline
KDE0v1 & \cite{Agrawal:2005ix} & 0.74 &  0.81&  0.165 &227  & 34.6  \\
LNS & \cite{cao06a}  &0.83 & 0.73 & 0.175 & 211  &  33.4  \\
NRAPRii& \cite{Steiner:2004fi}  &  0.69 & 0.60 & 0.161 & 226 & 32.8  \\
SQMC700 & \cite{Guichon:2006er} & 0.76 & 0.64 & 0.170 & 220 & 43.5  \\
SkM* & \cite{bartel82a}  &0.79 & 0.65 &0.160  & 217& 30.0  \\
SLy5 & \cite{chabanat98}  & 0.70 &0.80 & 0.160 & 230 &32.0  \\
T11 & \cite{Lesinski:2007zz}  & 0.70 & 0.80 & 0.161 &230 & 32.0  \\
T44 & \cite{Lesinski:2007zz} & 0.70 & 0.80  & 0.161 &230 &  32.0   \\
UNEDF0 & \cite{Kortelainen:2010}  & 1.11 & 0.80 & 0.161 & 230& 30.5\\
\hline
\end{tabular}
\caption{Basic SNM properties for the nine Skyrme
parametrizations used in the present work.}
\label{parametrizations}
\end{center}
\end{table}

Such a link can be established by means of random phase approximation (RPA) calculations of SNM. Within this approach, the instability of the density is marked by the occurrence of a pole at $\omega=0$ and finite momentum in the response function $\chi^{(\alpha)}_{\text{RPA}}(\vec{q}_{\text{ph}},\omega; k_{F})$  ($\alpha=(S,T)$), that is, by the existence of a zero-energy excitation mode. The solution of this implicit equation defines domains of instability in the $(\vec{q}_{\text{ph}},k_{F})$ plane whose boundaries define the curve $\rho_{p}(q_{\text{ph}})$, where $\rho \equiv 2 k^{3}_{\text{F}} / 3\pi^2$.
The formalism used to compute
$\chi^{(\alpha)}_{\text{RPA}}(\vec{q}_{\text{ph}},\omega; k_{F})$ on the basis of the Skyrme EDF has been outlined for central terms in
Ref.~\cite{garciarecio92a}, and extended to
spin-orbit~\cite{margueron06a} and tensor
terms~\cite{Davesne:2009ky,Pastore:2012mf}.
The extraction of unstable
modes at density $\rho$ and momentum transfer
$\vec{q}_{\text{ph}}$ can be efficiently performed on the basis of the inverse
energy-weighted sum-rule~\cite{Pastore:2012mf}.

The analysis is performed in three steps:
\begin{enumerate}
\item Each of the nine parametrizations is modified by increasing the value of $C^{\rho \Delta \rho}_{1}$ from its nominal value up to a critical value leading to non-convergence. In the following, the notation of a parametrization with a prime (as SLy5') indicates that the value of $C^{\rho \Delta \rho}_{1}$ is different from the one of the original parametrization. We have verified that the variation of $C^{\rho \Delta \rho}_{1}$ around its nominal value (in
steps of $0.1$ MeV fm$^{5}$) does not significantly deteriorate the overall properties of the parametrization.
\item The ground-state energy of selected nuclei is calculated with the modified parametrizations.  A parametrization should be discarded if it leads to non-convergence for any known nucleus. To make the calculation numerically manageable, we chose a representative set of nine spherical doubly closed-shell nuclei ($^{16}$O,$^{40,48}$Ca,$^{56,78}$Ni,$^{100,132,176}$Sn,$^{208}$Pb) and one deformed open-shell nucleus
($^{170}$Hf). Time-reversal symmetry can be enforced for  even-even nuclei, which suppresses possible spin ($S\!=\!1$) instabilities originating from time-odd terms of the EDF. Pairing correlations are also neglected. For each
of the nine parametrizations, we define the critical value $C^{\rho \Delta\rho}_{1,\text{crit}}$ as the smallest value of $C^{\rho \Delta \rho}_1$ for which an instability occurs in any of the selected nuclei.
\item For $C^{\rho \Delta\rho}_{1}=C^{\rho \Delta\rho}_{1,\text{crit}}$, we extract the minimum of $\rho_{p}(q_{\text{ph}})$, which we will refer to as $\rho_{\text{crit}}$ in the following, and study its correspondence to the results in finite nuclei.
\end{enumerate}

To facilitate the analysis of the results obtained in step 2 and 3 of our protocol, we introduce the scalar-isoscalar density at
the center of $^{40}$Ca $\rho_\mathrm{cent}=\rho_0(\vec{0})$ and the relative momentum distribution at $\vec{R}=\vec{0}$ through
\begin{equation}
f_i(\mathbf q)
=\frac{1}{(2\pi)^{3/2}}\int\!e^{-\mathrm{i}\mathbf q\cdot\mathbf s}
\rho_i(\vec{0},\mathbf s)\, d^3{s}\,,
\end{equation}
where $\rho_i(\mathbf R,\mathbf s)$ is the non-local scalar density for species
$i=n$ or $p$ as a function of the center of mass and relative coordinates
$\mathbf R$ and $\mathbf s$. We chose $^{40}$Ca because its density distribution
exhibits a central bump that can reach about $1.2\times\rho_{\text{sat}}$ and makes this nucleus 
particularly prone to instabilities (see below).

The convention used to define $f_i( q)$
is different of the one used in~\cite{casas} and allows to have
\begin{equation}
\rho_i(\vec{0},\vec{0})
=\frac{1}{(2\pi)^{3/2}}\int\!f_i(\mathbf q)\, d^3\vec{q}\,.
\end{equation}

\subsection{Numerical detections of instabilities}\label{subsection:protocol}

The detection of instabilities in nuclei depends on the numerical algorithm, on the symmetry restrictions imposed to solve the equations, and on the accuracy that is required in the calculation.
An insufficient accuracy or overly restrictive symmetries may hinder the instability and artificially increase the value of
$C^{\rho \Delta\rho}_{1,\text{crit}}$. To investigate such features and limit the numerical bias, we
employ two methods to perform the analysis in spherical symmetry and a third one allowing triaxial quadrupole deformations.
\begin{itemize}
\item
In \textsc{Hosphe}~\cite{Carlsson:2009mq}, single-particle wave
functions are expanded on an optimized spherical harmonic oscillator (HO) basis.
The accuracy is varied by changing the number of shells
$N_{\text{sh}}$ included in the HO basis.

\item
In \textsc{Lenteur}~\cite{lenteur}, one-body equations are solved
in coordinate space enforcing spherical symmetry. Radial wave functions with an angular momentum up to $41/2 \, \hbar$ are discretized on a mesh
along a radius of 18~fm. The accuracy of the calculation
is varied by modifying the step size $dr$ of the radial mesh.

\item
In \textsc{Ev8}~\cite{bonche05}, single-particle wave functions are
discretized on a 3-dimensional (3D) Cartesian mesh. Three symmetry planes are
imposed, corresponding to the description of triaxially deformed shapes.
The accuracy of the calculation is varied by changing the step size $dx$
of the mesh.

\end{itemize}

\section{Results}
\label{results}

\subsection{Analysis in nuclei}

Here, we carry out step 1 and 2 of the protocol described in Sect.~\ref{subsection:protocol}. In calculations of finite nuclei, as the one-body equations of motion are solved iteratively, an instability  in the $S=0$, $T=1$ channel occurs when it becomes energetically favorable to build oscillations of neutrons against protons of unlimited
amplitudes~\cite{lesinski06a}. Numerically,
this manifests itself differently in the various codes.
In \textsc{Hosphe}, the convergence criteria cannot be met beyond a certain value for $C^{\rho\Delta\rho}_{1}$, even when running the code for an unusually large number of iterations, and a random scatter of the energies is observed as a function of the iterations. For an insufficient number of shells, however, this is preceded by a very strong - unphysical - variation of the ground-state energy, indicating that the instability in fact occurs already at smaller values of $C^{\rho\Delta\rho}_{1}$  (see Appendix \ref{app:hosphe} for more details).
By contrast, with
\textsc{Lenteur} and \textsc{Ev8} an instability is clearly detected at all accuracies. For \textsc{Lenteur}, the calculation is halted
when density oscillations make the isovector gradient term
unreasonably large, whereas for \textsc{Ev8} the convergence criteria are never reached.

\begin{figure}[t!]
\includegraphics{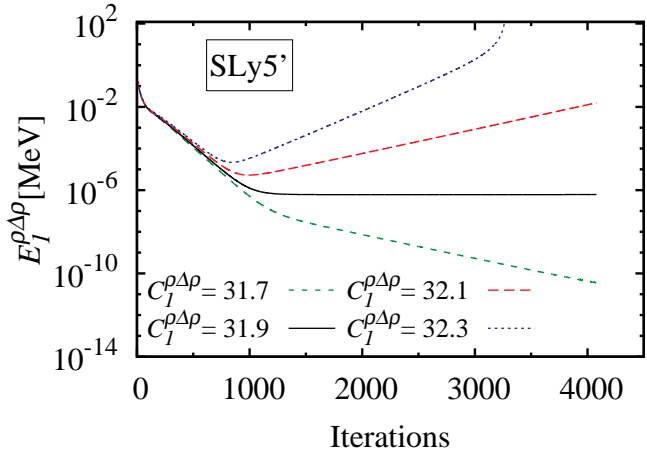}
\caption{
\label{SLY5ev8}
(Color online) Contribution of $E^{\rho\Delta\rho}_1$ to the binding energy of $^{40}$Ca as a function of the number of iterations. Four modified SLy5'~\cite{chabanat98} parametrizations with values of
$C^{\rho\Delta\rho}_1$ around its critical value $C^{\rho \Delta\rho}_{1,\text{crit}}$ are represented. Calculations are performed with the \textsc{Ev8} code for a value $dx=0.4$\,fm of the Cartesian mesh.}
\end{figure}

 Systematic exploratory calculations with both spherical codes showed that the instability always sets in first in $^{40}$Ca or $^{208}$Pb (see App.~\ref{app:spherical}). The calculations with the 3D code, being much more time-consuming at small mesh sizes, were
thus limited to these two doubly-magic nuclei and a deformed one, $^{170}$Hf. The values of $C^{\rho\Delta\rho}_{1,{\rm crit}}$ obtained with \textsc{Hosphe} for an unusually high number of shells, $N_{\text{sh}}$=60, are still significantly
larger than those obtained with methods using a coordinate space representation. Results with \textsc{Lenteur} and \textsc{Ev8} are consistent but $C^{\rho\Delta\rho}_{1,{\rm crit}}$ systematically takes lower values with \textsc{Ev8}. This is explained
by the lower degree of symmetry of the latter code which allows more freedom to develop oscillations in a ``spherical" nucleus. Also, with \textsc{Ev8}, the value of $C^{\rho\Delta\rho}_{1}$ at which the instability sets in is systematically the lowest in $^{40}$Ca. Therefore, in the following, we concentrate on the results obtained with \textsc{Ev8} for $^{40}$Ca. At $dx=0.40$ fm, the size of the box in all directions was chosen equal to 26 fm and for the larger values of $dx$ the number of points on the mesh was adjusted accordingly.

 We then benefit from $^{40}$Ca being a $N=Z$ nucleus and compute its energy without Coulomb interaction such that $E^{\rho\Delta\rho}_1$ should be zero at convergence. The calculation is initiated with wave functions of the $N\neq Z$ nucleus $^{50}$Ca such that $E^{\rho\Delta\rho}_1$ is non-zero when the iterations are started ($\sim10^{-1}$ MeV) and consequently
run for 4000 iterations.  Parametrizations are considered to be stable when the linear slope with which $E^{\rho\Delta\rho}_1$ changes in log scale after about 1000 iterations is negative (see
Fig.~\ref{SLY5ev8}) and $E^{\rho\Delta\rho}_1$ becomes of the order $10^{-10}$ MeV or less. This allows us to pin down $C^{\rho\Delta\rho}_{1,{\rm crit}}$ with a numerical uncertainty of about $0.2$ MeV fm$^{5}$ (see Table~\ref{cc} for $dx=0.4$\,fm).

\begin{table}
\begin{center}
\begin{tabular}{lcccccc}
\hline
Param. & Ref.
       & $C^{\rho \Delta \rho}_{1}$
       & $\rho_{\text{min}}/\rho_{\text{sat}}$
       & $C^{\rho \Delta \rho}_{1,\text{crit}}$
       & $\rho_{\text{crit}}/\rho_{\text{sat}}$  \\
\hline
KDE0v1  & \cite{Agrawal:2005ix}  & 11.498 & 2.39 &  30.8(1)  &    1.18 \\
LNS     & \cite{cao06a}          & 33.750 & 1.25 & 28.5(1)   &    1.35 \\
NRAPRii & \cite{Steiner:2004fi}  & 16.599 & 4.21 & 33.1(1)   &    1.67 \\
SQMC700 & \cite{Guichon:2006er}  & 15.884 & 4.77 & 31.1(1)   &    1.45 \\
SkM*    & \cite{bartel82a}       & 17.109 & 2.94 & 32.7(2)   &    1.36 \\
SLy5    & \cite{chabanat98}      & 16.375 & 1.72 & 31.7(2)   &    1.08 \\
T11     & \cite{Lesinski:2007zz} & 14.252 & 1.92 & 31.6(2)   &    1.08\\
T44     & \cite{Lesinski:2007zz} &  -4.300& 6.63  &  31.8(2) &    1.05 \\
UNEDF0 & \cite{Kortelainen:2010} &-55.623 & 4.13  & 29.0(1)  &    1.02\\
\hline
\end{tabular}
\caption{Nominal values of $C^{\rho \Delta \rho}_{1}$ (MeV\,fm$^{5}$) and $\rho_{\text{min}}/\rho_{\text{sat}}$. The critical coupling $C^{\rho \Delta \rho}_{1,\text{crit}}$ and density  $\rho_{\text{crit}}/\rho_{\text{sat}}$ are given for
$dx=0.4$\,fm (see text).}
\label{cc}
\end{center}
\end{table}

\begin{figure}[!t]
\includegraphics{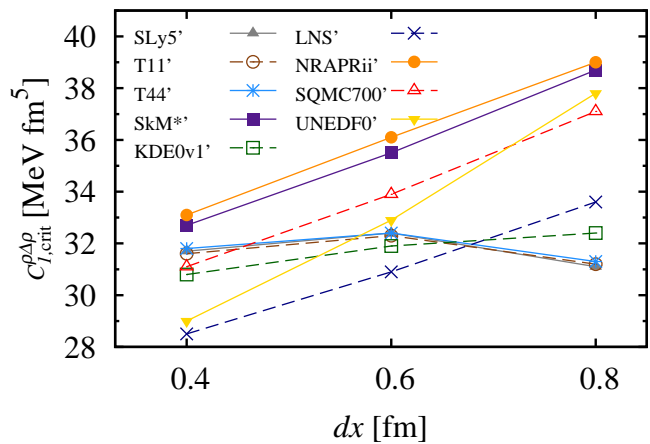}
\caption{(Color online)  Critical value of $C^{\rho \Delta\rho}_{1,\text{crit}}$ for the various (modified) parametrizations as a function of the step size $dx$ used in the \textsc{Ev8} code.}
\label{shellconv}
\end{figure}

A few comments are in order before the further presentation of our results. First, we have verified that pairing correlations do not alter the outcome of the analysis. Second, we note that the number of iterations necessary to unambiguously identify the instability is significantly larger than what is routinely used. It is thus easy to overlook the unstable nature of a given parametrization~\cite{lesinski06a}. The same is true when imposing too restrictive symmetries. As such, the values of $C^{\rho\Delta\rho}_{1,{\rm crit}}$ extracted with \textsc{Ev8} should be seen as upper bounds, as one cannot rule out that a completely symmetry-unrestricted numerical
representation would not provide even lower values. Last but not least, overly restrictive numerical parameters may also hide the instability as already mentioned for the code employing a HO basis expansion. The
sensitivity to the mesh size is illustrated for \textsc{Ev8} in Fig.~\ref{shellconv}, where the value of $C^{\rho\Delta\rho}_{1,{\rm crit}}$ is displayed for $dx=0.4, 0.6$ and $0.8$\,fm for the nine parametrizations studied here. One
observes a large change with $dx$ for LNS, NRAPRii, SQMC700, SkM*, and UNEDF0, whereas that variation is much milder for KDE0v1, SLy5, T11, and T44. While $C^{\rho\Delta\rho}_{1,{\rm crit}}$ varies over a range similar to its
numerical uncertainty for the latter group, it continues to decrease linearly for the former group as one lowers the mesh to $dx=0.4$\,fm, which is half of the value typically used in nuclear structure studies.

\begin{figure}[t!]
\includegraphics{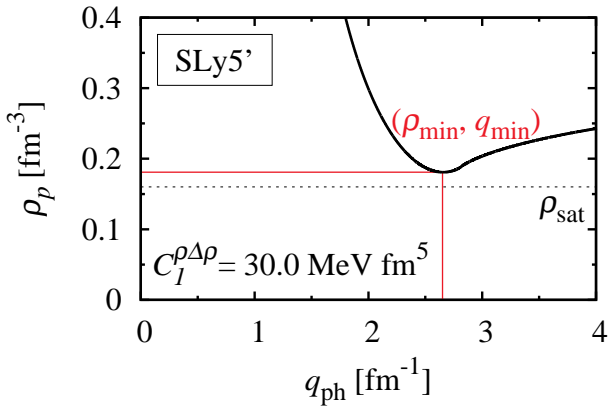}
\caption{
\label{lrfig1}
(Color online) The function $\rho_{p}(q_{\text{ph}})$ for
a SLy5' parametrization corresponding to $C^{\rho \Delta \rho}_{1}=30.0$ MeV fm$^5$. The minimum of the
$\rho_{p}(q_{\text{ph}})$ defines the lowest density $\rho_{\text{min}}$ at which a
pole occurs for this  value of $C^{\rho \Delta \rho}_{1}$.
}
\end{figure}

\subsection{Connection with RPA in SNM}

We now come to step 3 of our protocol (see Sect.~\ref{subsection:protocol}) and aim to establish a connection between the non-convergence occurring in calculations of finite nuclei with results obtained using the RPA in SNM.

\begin{figure}[t!]
\includegraphics{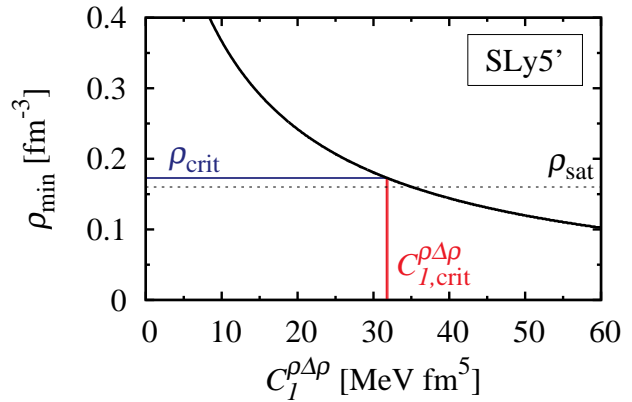}
\caption{
\label{lrfig}
(Color online)
$\rho_{\text{min}}$ as defined in Fig.~\ref{lrfig1} as a function of $C^{\rho \Delta \rho}_{1}$.
The vertical band $C^{\rho \Delta \rho}_{1,\text{crit}}$ intersects the
curve to define the horizontal band $\rho_{\text{crit}}$. The dashed line
denotes the saturation density $\rho_{\text{sat}}$ of SNM corresponding to by SLy5.}
\end{figure}

\begin{figure}
\includegraphics{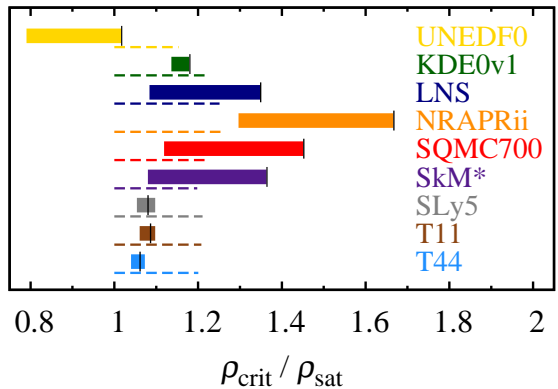}
\caption{(Color online)
Critical density $\rho_{\text{crit}}/\rho_{\text{sat}}$. The uncertainty band comes both from the numerical extraction and the  variation with $dx$. The vertical bar indicates the value for the lowest mesh size used
$dx=0.4$\,fm. The dashed line indicates the interval $[1,\rho_{\text{cent}}/\rho_{\text{sat}}]$.}
\label{bands}
\end{figure}

An example of an RPA calculation is presented in Fig.~\ref{lrfig1}, which displays $\rho_{p}(q_{\text{ph}})$ for SLy5' with $C^{\rho \Delta \rho}_{1}$ slightly below $C^{\rho \Delta \rho}_{1,{\rm crit}}$.  The value of $\rho_{\text{min}}$ is
defined as the minimum of $\rho_{p}(q_{\text{ph}})$ and corresponds to a momentum transfer $q_{\text{ph}}=q_{\text{min}}$. Fig.~\ref{lrfig}
presents the dependence of $\rho_{\text{min}}$ on $C^{\rho \Delta \rho}_{1}$. The critical density
$\rho_{\text{crit}}$ is then extracted as the value of $\rho_{\text{min}}$ obtained for $C^{\rho \Delta \rho}_{1}=C^{\rho \Delta \rho}_{1,\text{crit}}$. The values obtained for the nine EDF parametrizations are listed in Table~\ref{cc} and plotted in Fig.~\ref{bands}.
The uncertainty on $\rho_{\text{crit}}$ is estimated from the uncertainty on $C^{\rho \Delta \rho}_{1,\text{crit}}$
due to both its method of extraction and its dependence on $dx$. Figure~\ref{bands} also presents
the interval between $\rho_\mathrm{sat}$ and $\rho_\mathrm{cent}$, which is the highest density
attained in $^{40}$Ca. Note that
$\rho_\mathrm{cent}$ is typically about 20~\% larger than $\rho_\mathrm{sat}$.

Naively, one would expect to find a value of $\rho_{\text{crit}}$ at densities that are explored in a nucleus, otherwise stated, one would expect $\rho_{\text{crit}}\le\rho_{\text{cent}}$. This is the case for UNEDF0, KDE0v1, SLy5, T11 and T44 at all accuracies. By contrast, for LNS, SQMC700, and SkM* this is no longer true for all accuracies and for NRAPRii,  $\rho_{\text{crit}}$ at all accuracies even corresponds to densities that are never probed inside a nucleus. Clearly, the picture is more complicated than a naive one-to-one correspondence between the densities occurring in a finite nucleus and those probed by SNM, as could be expected when considering that the density at each point inside the nucleus is not behaving as if it simply were a piece of SNM with the same density.
Not surprisingly, the four parametrizations (KDE0v1, SLy5, T11 and T44) for which the value of $C^{\rho \Delta \rho}_{1,{\rm crit}}$ does not vary much with $dx$ (see Fig.~\ref{shellconv}), present much smaller error bars than five others.

\subsection{Discussion}

\begin{figure}[htbp]
\includegraphics{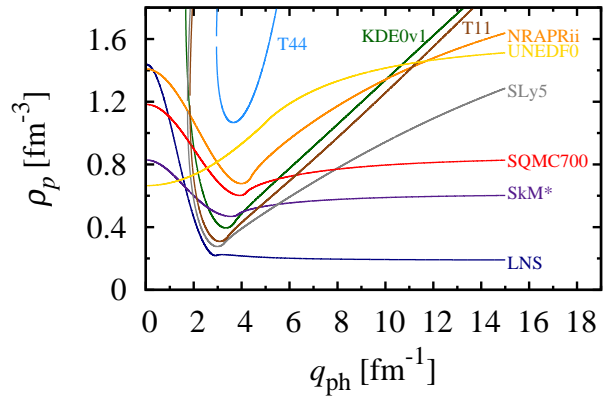}
\caption{(Color online) 
$\rho_{p}(q_{\text{ph}})$ for the parametrizations
given in Table~\ref{cc}  at the nominal value of the coupling constant $C_1^{\rho\Delta\rho}$.
}
\label{allint}
\end{figure}

\begin{figure}[htbp]
\includegraphics{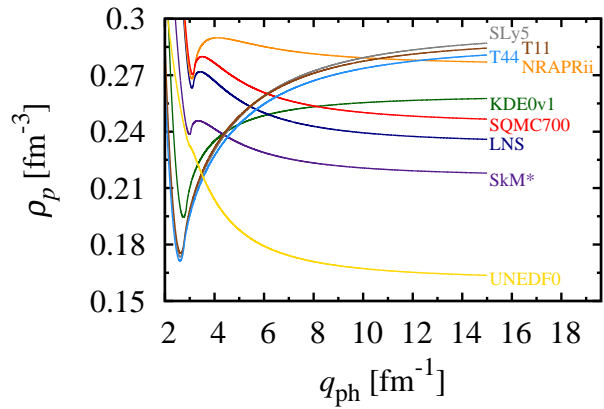}
\caption{(Color online) 
$\rho_{p}(q_{\text{ph}})$ for the parametrizations
given in Table~\ref{cc}  at the critical value of the coupling constant $C_1^{\rho\Delta\rho}$. Note the different scale of the y-axis in comparison to Fig. \ref{allint}.}
\label{allint2}
\end{figure}

Figure~\ref{allint} and Fig.~\ref{allint2}  provide the same kind of information as Fig.~\ref{lrfig1} for the nine parametrizations but extended to larger values of $q_{\text{ph}}$. In Fig.~\ref{allint}, $C^{\rho \Delta \rho}_{1}$ was taken at its nominal value. All curves present a well defined minimum at a small $q_{\text{ph}}$, except for LNS, for which $\rho_{p}(q_{\text{ph}})$ decreases monotonically above $q_{\text{ph}}=4$ fm$^{-1}$. This different behavior of the LNS curve can be attributed to the fact that LNS is already unstable for the nominal value of $C_1^{\rho\Delta\rho}$. In Fig.~\ref{allint2}, $C^{\rho \Delta \rho}_{1}$ was chosen at its critical value. Several differences between the two figures are clearly apparent. For all EDFs, the minimal value of the density for which an instability occurs at low $q_{\text{ph}}$ is significantly reduced. For the four parametrizations (KDE0v1, SLy5, T11 and T44) with small error bars in $\rho_{\text{crit}}/\rho_{\text{sat}}$, this minimal value corresponds to a well-defined minimum beyond which the curve increases in a monotonic way. By contrast, for the other five parametrizations (LNS, SQMC700, SkM*,UNEDF0), there is a monotonic decrease of the density corresponding to the pole for large values of $q_{\text{ph}}$ which seemingly approaches an asymptotic value.

\begin{figure}[htbp]
\includegraphics{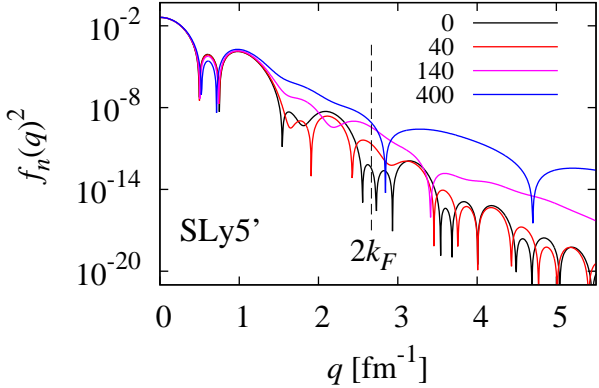}
\caption{(Color online) Square of the relative momentum distribution $f_{n}^{2}(q)$ at $R=0$ for neutrons
in ${}^{40}$Ca with SLy5' taking $C_1^{\rho\Delta\rho}$ slightly above $C_{1,\text{crit}}^{\rho\Delta\rho} $.
The four curves (see text) correspond to different numbers of Hartree-Fock iterations and the vertical
dashed line indicates $q=2k_F$.
}
\label{figfqa}
\end{figure}

\begin{figure}[htbp]
\includegraphics{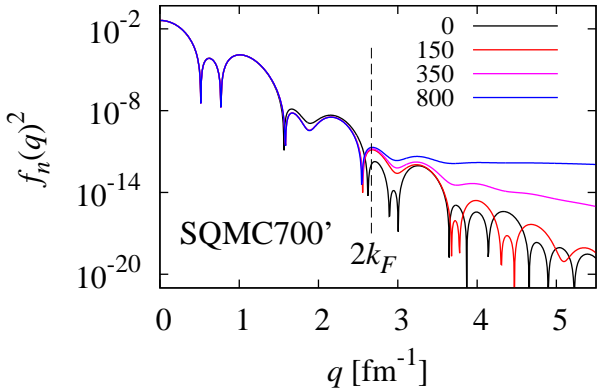}
\caption{(Color online) Same as Fig.~\ref{figfqa} but for SQMC700' ($C_1^{\rho\Delta\rho}$ slightly above $C_{1,\text{crit}}^{\rho\Delta\rho} $).}
\label{figfqb}
\end{figure}

Let us now look at the distribution of relative neutron momenta $f_{n}(q)$ at the center of $^{40}$Ca, i.e. where its density is maximal. The calculation is performed in spherical symmetry with the code \textsc{Lenteur}.  The starting point of the calculation is a converged wave function obtained for the nominal value $C_1^{\rho\Delta\rho}$ of a given parametrization. We then increase $C_1^{\rho\Delta\rho}$ to a value just above $C_{1,\mathrm{crit}}^{\rho\Delta\rho}$ and run the calculation for several hundreds of iterations. Figures~\ref{figfqa} and~\ref{figfqb} display $f_n^2(q)$ at various numbers of iterations on the way to the non-convergence for SLy5' and SQMC700', respectively.

Both parametrizations display a very different behavior. For SLy5', $f_n^2(q)$ starts to grow around $q = 2.2$~fm$^{-1}$ (see Fig.~\ref{allint}) and grows significantly for low $q$-values during the iterations (note the logarithmic scale). This indicates that the divergence is highly dominated by these low $q$-values and is consistent with the fact that $\rho_{p}(q_{\text{ph}})$ exhibits a clear global minimum at small $q_{\text{ph}}$ for SLy5'. For SQMC700', there is only an increase of  $f_n^2(q)$ for high $q$-values. Indeed, when the calculation starts to diverge, all the weights of momenta larger than $q = 3.1$~fm$^{-1}$ increase. This seems to be consistent with $\rho_{p}(q_{\text{ph}})$ being a monotonously decreasing function for $q_{\text{ph}}$ tending to $\infty$, without a distinct global minimum. 

The observations made above for SLy5' (SQMC700') are valid for any parametrization belonging to the first (second) group. Hence, the very different behavior of $f_n^2(q)$ at the onset of and during the divergence seems to support the hypothesis that there is a link between the $q_{\text{min}}$ in SNM and the onset of instabilities in finite nuclei.

\section{Conclusions}
\label{conclusions}

The present study aimed at relating instabilities in energy density functional calculations of
nuclei to finite-wavelength instabilities of homogeneous symmetric nuclear matter computed at the RPA level. A detailed study of the
various numerical aspects in finite nuclei and of the relation between results in SNM and finite nuclei has
lead us to the following conclusions:
\begin{enumerate}
\item
Instabilities of finite nuclei can be artificially hidden when using
inappropriate numerical schemes, such as an insufficiently large basis,
a too coarse mesh, or overly restrictive, e.g.\ spherical,
symmetries. An unusually high accuracy of the calculation is required to
unambiguously detect instabilities.
Also, a code breaking spherical symmetry and relying on the discretization
of a 3D mesh seems to be a more appropriate tool for this task
\item
Numerically hiding a finite-size instability by choosing too coarse a
numerical representation is not equivalent to suppressing it. Although
a converged solution of the self-consistent mean-field equations
might be found for a given set of numerical parameters, the observables
remain strongly dependent on the choices made for these parameters.
\item Omitting the UNEDF0 parametrization for the moment, the parametrizations that we have studied can be systematically classified in two groups. The first one (SLy5, KDE0v1, T11, T44) corresponds to parametrizations for which the uncertainty in
$C^{\rho\Delta\rho}_{1,{\rm crit}}$ is small, $\rho_{p}(q_{\text{ph}})$ has a marked absolute minimum as a function of $q_{\text{ph}}$, and the behavior of $f_{n}^{2}(q)$ indicates that the divergence is dominated by low $q$-values. For this group of parametrizations, $\rho_{\text{crit}}$ within error bars is smaller than $\rho_{\text{cent}}$. In case of the second group (LNS, NRAPRii, SQMC700, SkM*), the uncertainty in $C^{\rho\Delta\rho}_{1,{\rm crit}}$ is large, $\rho_{p}(q_{\text{ph}})$ is monotonously decreasing beyond $q_{\text{ph}}\approx 3$ fm$^{-1}$, and $f_{n}^{2}(q)$ grows predominantly for large $q$-values when the calculations diverge.

\item The same grouping of the parametrizations is observed when considering the quality of their prediction of nuclear masses. SLy5, T44, T11, and KDE0v1 were adjusted treating properties of nuclear matter and finite nuclei on similar footing. By contrast, no data on nuclear masses were included in the protocol for LNS and SQMC700.  As already pointed out in the original references,  the mass residuals for these parametrizations are prohibitively large ($ > 5 \%$). NRAPR was adjusted in a similar manner, with only the spin-orbit part of the EDF being tuned to the binding energies of $^{208}$Pb, $^{90}$Zr and $^{40}$Ca, but with very poor results. Finally, SkM$^*$ has been constructed using a semi-classical approximation of the mean-field and also leads to non-satisfactory results for binding energies.
\item Finally, let us consider UNEDF0. It has a large uncertainty in $C^{\rho\Delta\rho}_{1,{\rm crit}}$ and a $\rho_{p}(q_{\text{ph}})$ that decreases systematically with $q_{\text{ph}}$, hence placing it in the second group of parametrizations. However,  $\rho_{\text{crit}}<\rho_{\text{cent}}$ and properties of nuclear matter and finite nuclei were treated on equal footing during its adjustment, thus associating it with the first group of parametrizations. We do not see a clear explanation of this mixed behavior. One can however note that the value found for $C^{\rho\Delta\rho}_{1,{\rm crit}}$ is far from its nominal value, even if it falls within the error bars reported in \cite{Kortelainen:2010} .
\end{enumerate}

Combining all aspects of the study carried out in the present article, no universal quantitative picture emerges regarding the value of $\rho_{\text{crit}}$ when scanning various Skyrme parametrizations : Omitting UNEDF0, which has a large uncertainty in $\rho_{\text{crit}}$ and for which $C^{\rho\Delta\rho}_{1,{\rm crit}}$ is far from its nominal value, one finds a group of parameterizations (SLy5, KDE0v1, T11, T44)  that delivers a high quality prediction of nuclear masses and for which the $\rho_{\text{crit}}$ determined by means of RPA for SNM corresponds to densities that are probed in a finite nucleus. On the other hand, one finds a group of parameterizations (LNS, NRAPRii, SQMC700, SkM*) which leads to less satisfactory results in the description of nuclear masses and for which there seems to be no easy one-to-one correspondence between $\rho_{\text{crit}}$ and the densities probed in a nucleus. Note, however, that LNS, NRAPRii and SQMC700 are shown to be consistent with a large set of pseudo-data in SNM.

For the purpose of constructing parametrizations for the description of finite nuclei that are stable with respect to scalar-isovector perturbations, we propose a two-fold criterion: First the minimum of $\rho_{p}(q_{\text{ph}})$ should be larger than the central density in $^{40}$Ca, in practice around $1.2$ times the saturation density. In addition, one has also to verify that $\rho_{p}(q_{\text{ph}})$ exhibits a distinct global minimum and is not a monotonously decreasing function for large transferred momenta.

Since RPA calculations of SNM can be performed at no computational cost, the above stability criterion can be easily incorporated in fitting protocols to identify and reject (near-) unstable regions of the parameter space when adjusting the coefficients of the EDF. The value of this threshold should of course be increased in case one is interested in nuclear systems exploring densities higher than those encountered in nuclear ground states, such as they appear in neutron stars for example.

\section{Acknowledgements}
M.B.\ and K.B.\ thank P.-G.\ Reinhard for a clarifying discussion
about the RPA in infinite matter.
This work was supported by the Agence
Nationale de la Recherche under Grant No.~ANR 2010 BLANC 0407 "NESQ",
by the CNRS/IN2P3 through the PICS No.~5994, by the European Union's Seventh Framework Programme ENSAR under grant agreement n°262010 and by the Belgian Office for Scientific Policy under Grant No. PAI-P7-12. V.H. acknowledges financial support from the F.R.S.-FNRS Belgium.

\appendix

\section{Observation of finite-size instabilities with \textsc{Hosphe}}\label{app:hosphe}

\begin{figure}[!tb]
\begin{center}
\includegraphics[width=8cm]{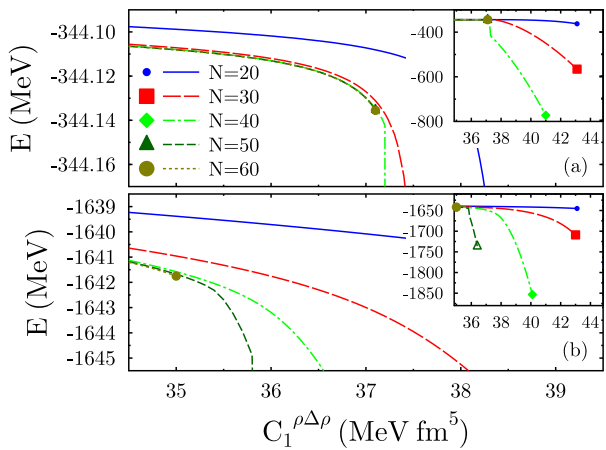}
\end{center}
\caption{
\label{shellSLY5}
(Color online) Binding energy of (a) $^{40}$Ca and (b) $^{208}$Pb
obtained from \textsc{Hosphe} and SLy5~\cite{chabanat98} as a function
of $C^{\rho \Delta \rho}_{1}$.
Calculations are performed for several
values of the number of shells $N$. The curves end at the largest
value of $C^{\rho \Delta \rho}_{1}$ for which the convergence criteria
are reached within 40 000 iterations.
}
\end{figure}

\begin{figure}[!tb]
\begin{center}
\includegraphics[width=0.35\textwidth,angle=-90]{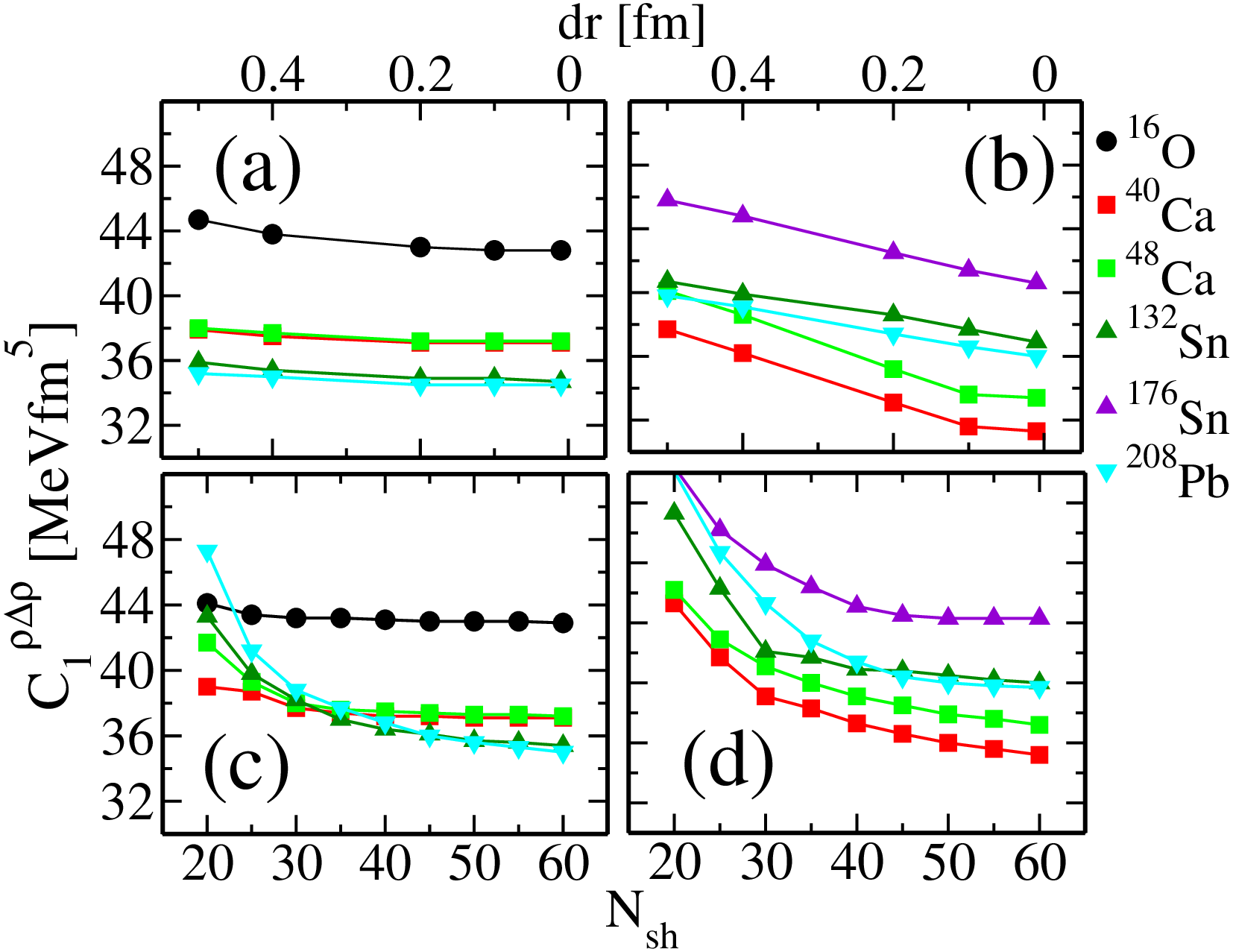}
\end{center}
\caption{\label{convspherical}
(Color online)
Maximum value of $C^{\rho \Delta \rho}_{1}$
for which a solution is found for the ground state of a given nucleus.
Results are displayed for SLy5 ((a) and (c)) and T44 ((b) and (d)). Results are
displayed in the upper (lower) panel as a function of the mesh (basis)
size used in \textsc{Lenteur} (\textsc{Hosphe}).
}
\end{figure}

In a code based on an expansion on an oscillator basis such as \textsc{Hosphe}, the occurrence of instabilities
manifests itself differently than in codes based on a discretization on a 1D or a 3D mesh. This is illustrated in
Fig.~\ref{shellSLY5} for the \mbox{$N=Z$} nucleus~$^{40}$Ca and for
$^{208}$Pb. The binding energy is plotted as a function of $C^{\rho \Delta \rho}_{1}$ for different values of $N_{\text{sh}}$.
For both nuclei, the energy decreases up to a value of $C^{\rho \Delta \rho}_{1}$ where the convergence criteria on energy cannot
be met in spite of an unusually large number of iterations (up to 40000 in our calculations).
For the smaller number of shells shown in the Figure, ($N_{\text{sh}}\le 40$ for $^{40}$Ca
and $N_{\text{sh}}\le 50$ for $^{208}$Pb), an apparent convergence is obtained for values of $C^{\rho \Delta \rho}_{1}$ in the vicinity of $C^{\rho \Delta \rho}_{1,\text{crit}}$ but with
an unphysical large value of the binding energy. In that region of the plot, a 1 MeV fm$^{5}$ increase of $C^{\rho \Delta \rho}_{1}$ leads to a change of total energy of
the order of tens of keV for $N_{\text{sh}}=20$ and of tens of MeV for $N_{\text{sh}}=30,40$
in the case of $^{40}$Ca. These numbers have to be compared with those of \textsc{Ev8}. Then, for all mesh sizes, the energy only varies by a few keV for a 1 MeV fm$^{5}$ step in $C^{\rho \Delta \rho}_{1}$ below $C^{\rho \Delta \rho}_{1,\text{crit}}$. Increasing $N_{\text{sh}}$ to unusually large values
makes the detection of instabilities easier, but still leads with 60 shells to values of $C^{\rho \Delta \rho}_{1}$ significantly larger than with codes
based on an discretization on a mesh.
 This result clearly illustrates the impossibility
of an accurate determination of the value of $C^{\rho \Delta \rho}_{1}$ for which
the instability sets in when an expansion on an oscillator basis is used.
It also demonstrates that if an EDF parameterization is unstable, the occurrence of finite-size instabilities can be obscured
by a choice of a number of oscillator shells leading to an apparent convergence of binding energies but artificially suppressing the occurrence of instabilities.

\section{Results from the spherical codes}\label{app:spherical}

Systematic calculations for $^{16}$O, $^{40,48}$Ca, $^{56,78}$Ni, $^{100,132,176}$Sn, $^{208}$Pb
with both \textsc{Hosphe} and \textsc{Lenteur} have
shown that the lowest $C^{\rho \Delta \rho}_{1}$ at which the instability sets in
is found for $^{40}$Ca or $^{208}$Pb. This is illustrated in
Figure~\ref{convspherical} for the SLy5 and T44 parameterizations. The critical value of
$C^{\rho \Delta \rho}_{1}$  is obtained in $^{40}$Ca for T44 and in
$^{208}$Pb for SLy5. In the latter case, several nuclei lead to similar
values. Throughout our analysis and with the highest accuracy used in the
two spherical codes, the lowest values of $C^{\rho \Delta \rho}_{1,\text{crit}}$  are
systematically obtained with \textsc{Lenteur}.


\begin{thebibliography}{28}
\expandafter\ifx\csname natexlab\endcsname\relax\def\natexlab#1{#1}\fi
\expandafter\ifx\csname bibnamefont\endcsname\relax
  \def\bibnamefont#1{#1}\fi
\expandafter\ifx\csname bibfnamefont\endcsname\relax
  \def\bibfnamefont#1{#1}\fi
\expandafter\ifx\csname citenamefont\endcsname\relax
  \def\citenamefont#1{#1}\fi
\expandafter\ifx\csname url\endcsname\relax
  \def\url#1{\texttt{#1}}\fi
\expandafter\ifx\csname urlprefix\endcsname\relax\def\urlprefix{URL }\fi
\providecommand{\bibinfo}[2]{#2}
\providecommand{\eprint}[2][]{\url{#2}}

\bibitem[{\citenamefont{Bender et~al.}(2003)\citenamefont{Bender, Heenen, and
  Reinhard}}]{bender03b}
\bibinfo{author}{\bibfnamefont{M.}~\bibnamefont{Bender}},
  \bibinfo{author}{\bibfnamefont{P.-H.} \bibnamefont{Heenen}},
  \bibfont{and} \bibinfo{author}{\bibfnamefont{P.-G.}
  \bibnamefont{Reinhard}}, \bibinfo{journal}{Rev. Mod. Phys.}
  \textbf{\bibinfo{volume}{75}}, \bibinfo{pages}{121} (\bibinfo{year}{2003}).
\bibitem[{\citenamefont{Stringari and
  Treiner}(1987{\natexlab{a}})}]{stringari87b}
  \bibinfo{author}{\bibfnamefont{S.}~\bibnamefont{Stringari}} \bibnamefont{and}
  \bibinfo{author}{\bibfnamefont{J.}~\bibnamefont{Treiner}},
  \bibinfo{journal}{Phys. Rev. B} \textbf{\bibinfo{volume}{36}},
  \bibinfo{pages}{8369} (\bibinfo{year}{1987}{\natexlab{a}}).
\bibitem{barranco06a}
M. Barranco, R. Guardiola, S. Herna{\'n}dez, R. Mayol,
J. Navarro, and M. Pi,
J. Low Temp. Phys. {\bf 142}, 1 (2006).
\bibitem[{\citenamefont{Stringari and
  Treiner}(1987{\natexlab{b}})}]{stringari87a}
\bibinfo{author}{\bibfnamefont{S.}~\bibnamefont{Stringari}} \bibnamefont{and}
  \bibinfo{author}{\bibfnamefont{J.}~\bibnamefont{Treiner}},
  \bibinfo{journal}{J. Chem. Phys. A} \textbf{\bibinfo{volume}{87}},
  \bibinfo{pages}{5021} (\bibinfo{year}{1987}{\natexlab{b}}).
\bibitem{weisgerber92a}
S. Weisgerber, P.-G. Reinhard,
Z. Phys. D {\bf 23}, 275 (1992).

\bibitem{Bulgac:2008tm}
A. Bulgac, M. Forbes,
Phys. Rev. Lett. {\bf 101}, 215301 (2008).

\bibitem[{\citenamefont{Tondeur et~al.}(1984)\citenamefont{Tondeur, Brack,
  Farine, and Pearson}}]{tondeur84}
\bibinfo{author}{\bibfnamefont{F.}~\bibnamefont{Tondeur}},
  \bibinfo{author}{\bibfnamefont{M.}~\bibnamefont{Brack}},
  \bibinfo{author}{\bibfnamefont{M.}~\bibnamefont{Farine}}, \bibnamefont{and}
  \bibinfo{author}{\bibfnamefont{J.~M.} \bibnamefont{Pearson}},
  \bibinfo{journal}{Nucl. Phys. A} \textbf{\bibinfo{volume}{420}},
  \bibinfo{pages}{297} (\bibinfo{year}{1984}).
\bibitem[{\citenamefont{Lesinski et~al.}(2006)\citenamefont{Lesinski,
  Bennaceur, Duguet, and Meyer}}]{lesinski06a}
\bibinfo{author}{\bibfnamefont{T.}~\bibnamefont{Lesinski}},
  \bibinfo{author}{\bibfnamefont{K.}~\bibnamefont{Bennaceur}},
  \bibinfo{author}{\bibfnamefont{T.}~\bibnamefont{Duguet}}, \bibnamefont{and}
  \bibinfo{author}{\bibfnamefont{J.}~\bibnamefont{Meyer}},
  \bibinfo{journal}{Phys. Rev. C} \textbf{\bibinfo{volume}{74}},
  \bibinfo{pages}{044315} (\bibinfo{year}{2006}).
\bibitem{schunck10b}
N. Schunck, J. Dobaczewski, J. McDonnell, J. Mor{\'e}, W. Nazarewicz,
J. Sarich, M.~V. Stoitsov, Phys. Rev. C {\bf 81}, 024316 (2010).
\bibitem[{\citenamefont{Hellemans
  et~al.}(2012{\natexlab{b}})\citenamefont{Hellemans, Heenen, and
  Bender}}]{Hellemans:2011aa}
\bibinfo{author}{\bibfnamefont{V.}~\bibnamefont{Hellemans}},
  \bibinfo{author}{\bibfnamefont{P.-H.}~\bibnamefont{Heenen}}, \bibnamefont{and}
  \bibinfo{author}{\bibfnamefont{M.}~\bibnamefont{Bender}},
  \bibinfo{journal}{Phys. Rev. C} \textbf{\bibinfo{volume}{85}},
  \bibinfo{pages}{014326} (\bibinfo{year}{2012}{\natexlab{b}}).
\bibitem[{\citenamefont{Midgal}(1967)}]{migdal67a}
\bibinfo{author}{\bibfnamefont{A.~B.} \bibnamefont{Midgal}},
  \emph{\bibinfo{title}{Theory of Finite Fermi Systems and Applications to
  Atomic Nuclei}} (\bibinfo{publisher}{Wiley}, \bibinfo{address}{New York},
  \bibinfo{year}{1967}).
\bibitem{navarro13}{
J. Navarro and A. Polls,
Phys. Rev. C \textbf{87}, 044329 (2013).
}  
\bibitem[{\citenamefont{Vidana and Bombaci}(2002)}]{Vidana:2002pc}
\bibinfo{author}{\bibfnamefont{I.}~\bibnamefont{Vida\~na}} \bibnamefont{and}
  \bibinfo{author}{\bibfnamefont{I.}~\bibnamefont{Bombaci}},
  \bibinfo{journal}{Phys. Rev. C} \textbf{\bibinfo{volume}{66}},
  \bibinfo{pages}{045801} (\bibinfo{year}{2002}).
\bibitem[{\citenamefont{Bombaci et~al.}(2006)\citenamefont{Bombaci, Polls,
  Ramos, Rios, and Vidana}}]{Bombaci:2005vi}
\bibinfo{author}{\bibfnamefont{I.}~\bibnamefont{Bombaci}},
  \bibinfo{author}{\bibfnamefont{A.}~\bibnamefont{Polls}},
  \bibinfo{author}{\bibfnamefont{A.}~\bibnamefont{Ramos}},
  \bibinfo{author}{\bibfnamefont{A.}~\bibnamefont{Rios}}, \bibnamefont{and}
  \bibinfo{author}{\bibfnamefont{I.}~\bibnamefont{Vida\~na}},
  \bibinfo{journal}{Phys. Lett. B} \textbf{\bibinfo{volume}{632}},
  \bibinfo{pages}{638} (\bibinfo{year}{2006}).

\bibitem[{\citenamefont{Garc\'{i}a-Recio
  et~al.}(1992)\citenamefont{Garc\'{i}a-Recio, Navarro, Giai, and
  Salcedo}}]{garciarecio92a}
\bibinfo{author}{\bibfnamefont{C.}~\bibnamefont{Garc\'{i}a-Recio}},
  \bibinfo{author}{\bibfnamefont{J.}~\bibnamefont{Navarro}},
  \bibinfo{author}{\bibnamefont{Nguyen Van Giai}}, \bibnamefont{and}
  \bibinfo{author}{\bibfnamefont{N.~N.} \bibnamefont{Salcedo}},
  \bibinfo{journal}{Ann. Phys.} \textbf{\bibinfo{volume}{214}},
  \bibinfo{pages}{293} (\bibinfo{year}{1992}).
\bibitem[{\citenamefont{Margueron et~al.}(2006)\citenamefont{Margueron,
  Navarro, and Giai}}]{margueron06a}
\bibinfo{author}{\bibfnamefont{J.}~\bibnamefont{Margueron}},
  \bibinfo{author}{\bibfnamefont{N.}~\bibnamefont{Van Giai}}, \bibnamefont{and}
  \bibinfo{author}{\bibnamefont{J. Navarro}},
  \bibinfo{journal}{Phys. Rev. C} \textbf{\bibinfo{volume}{74}},
  \bibinfo{pages}{015805} (\bibinfo{year}{2006}).
\bibitem[{\citenamefont{Davesne et~al.}(2009)\citenamefont{Davesne, Martini,
  Bennaceur, and Meyer}}]{Davesne:2009ky}
\bibinfo{author}{\bibfnamefont{D.}~\bibnamefont{Davesne}},
  \bibinfo{author}{\bibfnamefont{M.}~\bibnamefont{Martini}},
  \bibinfo{author}{\bibfnamefont{K.}~\bibnamefont{Bennaceur}},
  \bibnamefont{and} \bibinfo{author}{\bibfnamefont{J.}~\bibnamefont{Meyer}},
  \bibinfo{journal}{Phys. Rev. C} \textbf{\bibinfo{volume}{80}},
  \bibinfo{pages}{024314} (\bibinfo{year}{2009});
\bibinfo{journal}{ibid.} \textbf{\bibinfo{volume}{84}},
  \bibinfo{pages}{059904(E)} (\bibinfo{year}{2011}).
\bibitem[{\citenamefont{Pastore et~al.}(2012)\citenamefont{Pastore, Davesne,
  Lallouet, Martini, Bennaceur et~al.}}]{Pastore:2012mf}
\bibinfo{author}{\bibfnamefont{A.}~\bibnamefont{Pastore}},
  \bibinfo{author}{\bibfnamefont{D.}~\bibnamefont{Davesne}},
  \bibinfo{author}{\bibfnamefont{Y.}~\bibnamefont{Lallouet}},
  \bibinfo{author}{\bibfnamefont{M.}~\bibnamefont{Martini}},
  \bibinfo{author}{\bibfnamefont{K.}~\bibnamefont{Bennaceur}},
  \bibnamefont{and} \bibinfo{author}{\bibfnamefont{J.}~\bibnamefont{Meyer}},
  \bibinfo{journal}{Phys. Rev. C} \textbf{\bibinfo{volume}{85}},
  \bibinfo{pages}{054317} (\bibinfo{year}{2012}).


\bibitem{ducoin08a}
C. Ducoin, J. Margueron, Ph. Chomaz,
Nucl. Phys. A {\bf 809}, 30 (2008).


\bibitem{ducoin08b}
C. Ducoin, C. Provid\^encia, A.M. Santos, L. Brito, Ph. Chomaz,
Phys. Rev. C {\bf 78}, 055801 (2008).

\bibitem{ducoin07}
C. Ducoin, Ph. Chomaz, F. Gulminelli,
Nucl. Phys. A {\bf 789}, 403 (2007).

\bibitem[{\citenamefont{Kortelainen and Lesinski}(2010)}]{Kortelainen:2010wb}
\bibinfo{author}{\bibfnamefont{M.}~\bibnamefont{Kortelainen}} \bibnamefont{and}
  \bibinfo{author}{\bibfnamefont{T.}~\bibnamefont{Lesinski}},
  \bibinfo{journal}{J. Phys. G} \textbf{\bibinfo{volume}{37}},
  \bibinfo{pages}{064039} (\bibinfo{year}{2010}).

\bibitem{sulaksano11a}
A. Sulaksono, Kasmudin, T.~J. B{\"u}rvenich, P.-G. Reinhard, J.~A. Maruhn,
Int. J. Mod. Phys. E {\bf 20}, 81 (2011).

\bibitem{pototzky10a}
K. J. Pototzky, J. Erler, P.-G. Reinhard, V. O. Nesterenko,
 Eur. J. Phys A \textbf{46}, 299 (2010).

\bibitem{stevenson12x}
P.~D. Stevenson, P.~M. Goddard, J.~R. Stone, M. Dutra,
Proceedings of XXXV Reuni{\~a}o de Trabalho sobre F{\'i}sica
Nuclear no Brasil, held at S{\~a}o Sebasti{\~a}o, S{\~a}o Paulo, Brazil, 2Ð6 September 2012,
F. L. Melquiades, F. A. Genezini, N. H. Medina , R. M. dos Anjos and S. dos Santos Avancini [edts.]
AIP Conf. Proc.  \textbf{1529}, 262 (2013).

\bibitem[{\citenamefont{Dutra et~al.}(2012)\citenamefont{Dutra, Lourenco,
  Sa~Martins, Delfino, Stone, and Stevenson}}]{Dutra:2012mb}
\bibinfo{author}{\bibfnamefont{M.}~\bibnamefont{Dutra}},
  \bibinfo{author}{\bibfnamefont{O.}~\bibnamefont{Louren{\c c}o}},
  \bibinfo{author}{\bibfnamefont{J.~S.}~\bibnamefont{S\'a~Martins}},
  \bibinfo{author}{\bibfnamefont{A.}~\bibnamefont{Delfino}},
  \bibinfo{author}{\bibfnamefont{J.~R.}~\bibnamefont{Stone}}, \bibnamefont{and}
  \bibinfo{author}{\bibfnamefont{P.~D.}~\bibnamefont{Stevenson}},
  \bibinfo{journal}{Phys. Rev. C} \textbf{\bibinfo{volume}{85}},
  \bibinfo{pages}{035201} (\bibinfo{year}{2012}).

\bibitem[{\citenamefont{Agrawal et~al.}(2005)\citenamefont{Agrawal, Shlomo, and
  Au}}]{Agrawal:2005ix}
\bibinfo{author}{\bibfnamefont{B.~K.}~\bibnamefont{Agrawal}},
  \bibinfo{author}{\bibfnamefont{S.}~\bibnamefont{Shlomo}}, \bibnamefont{and}
  \bibinfo{author}{\bibfnamefont{V.~K.} \bibnamefont{Au}},
  \bibinfo{journal}{Phys. Rev. C} \textbf{\bibinfo{volume}{72}},
  \bibinfo{pages}{014310} (\bibinfo{year}{2005}).

\bibitem[{\citenamefont{Cao et~al.}(2006)\citenamefont{Cao, Lombardo, Shen, and
  N.~V.~Giai}}]{cao06a}
\bibinfo{author}{\bibfnamefont{L.~G.} \bibnamefont{Cao}},
  \bibinfo{author}{\bibfnamefont{U.}~\bibnamefont{Lombardo}},
  \bibinfo{author}{\bibfnamefont{C.~W.} \bibnamefont{Shen}},
  \bibnamefont{and} \bibinfo{author}{\bibnamefont{Nguyen Van Giai}},
  \bibinfo{journal}{Phys. Rev. C} \textbf{\bibinfo{volume}{73}},
  \bibinfo{pages}{014313} (\bibinfo{year}{2006}).

\bibitem[{\citenamefont{Steiner et~al.}(2005)\citenamefont{Steiner, Prakash,
  Lattimer, and Ellis}}]{Steiner:2004fi}
\bibinfo{author}{\bibfnamefont{A.~W.} \bibnamefont{Steiner}},
  \bibinfo{author}{\bibfnamefont{M.}~\bibnamefont{Prakash}},
  \bibinfo{author}{\bibfnamefont{J.~M.} \bibnamefont{Lattimer}},
  \bibnamefont{and} \bibinfo{author}{\bibfnamefont{P.~J.} \bibnamefont{Ellis}},
  \bibinfo{journal}{Phys. Rep.} \textbf{\bibinfo{volume}{411}},
  \bibinfo{pages}{325} (\bibinfo{year}{2005}).

\bibitem[{\citenamefont{Guichon et~al.}(2006)\citenamefont{Guichon, Matevosyan,
  Sandulescu, and Thomas}}]{Guichon:2006er}
\bibinfo{author}{\bibfnamefont{P.~A.~M.} \bibnamefont{Guichon}},
  \bibinfo{author}{\bibfnamefont{H.~H.} \bibnamefont{Matevosyan}},
  \bibinfo{author}{\bibfnamefont{N.}~\bibnamefont{Sandulescu}},
  \bibnamefont{and} \bibinfo{author}{\bibfnamefont{A.~W.}
  \bibnamefont{Thomas}}, \bibinfo{journal}{Nucl. Phys. A}
  \textbf{\bibinfo{volume}{772}}, \bibinfo{pages}{1} (\bibinfo{year}{2006}).

\bibitem[{\citenamefont{Bartel et~al.}(1982)\citenamefont{Bartel, Quentin,
  Brack, Guet, and H{\aa}kansson}}]{bartel82a}
\bibinfo{author}{\bibfnamefont{J.}~\bibnamefont{Bartel}},
  \bibinfo{author}{\bibfnamefont{P.}~\bibnamefont{Quentin}},
  \bibinfo{author}{\bibfnamefont{M.}~\bibnamefont{Brack}},
  \bibinfo{author}{\bibfnamefont{C.}~\bibnamefont{Guet}}, \bibnamefont{and}
  \bibinfo{author}{\bibfnamefont{H.-B.} \bibnamefont{H{\aa}kansson}},
  \bibinfo{journal}{Nucl. Phys. A} \textbf{\bibinfo{volume}{386}},
  \bibinfo{pages}{79} (\bibinfo{year}{1982}).

\bibitem[{\citenamefont{Chabanat et~al.}(1998)\citenamefont{Chabanat, Bonche,
  Haensel, Meyer, and Schaeffer}}]{chabanat98}
\bibinfo{author}{\bibfnamefont{E.}~\bibnamefont{Chabanat}},
  \bibinfo{author}{\bibfnamefont{P.}~\bibnamefont{Bonche}},
  \bibinfo{author}{\bibfnamefont{P.}~\bibnamefont{Haensel}},
  \bibinfo{author}{\bibfnamefont{J.}~\bibnamefont{Meyer}}, \bibnamefont{and}
  \bibinfo{author}{\bibfnamefont{R.}~\bibnamefont{Schaeffer}},
  \bibinfo{journal}{Nucl. Phys. A} \textbf{\bibinfo{volume}{635}},
  \bibinfo{pages}{231} (\bibinfo{year}{1998}).

\bibitem[{\citenamefont{Lesinski et~al.}(2007)\citenamefont{Lesinski, Bender,
  Bennaceur, Duguet, and Meyer}}]{Lesinski:2007zz}
\bibinfo{author}{\bibfnamefont{T.}~\bibnamefont{Lesinski}},
  \bibinfo{author}{\bibfnamefont{M.}~\bibnamefont{Bender}},
  \bibinfo{author}{\bibfnamefont{K.}~\bibnamefont{Bennaceur}},
  \bibinfo{author}{\bibfnamefont{T.}~\bibnamefont{Duguet}}, \bibnamefont{and}
  \bibinfo{author}{\bibfnamefont{J.}~\bibnamefont{Meyer}},
  \bibinfo{journal}{Phys. Rev. C} \textbf{\bibinfo{volume}{76}},
  \bibinfo{pages}{014312} (\bibinfo{year}{2007}).

\bibitem{Kortelainen:2010}
M. Kortelainen, T. Lesinski, J. Mor\'e, W. Nazarewicz, J. Sarich, N. Schunck, M. V. Stoitsov and S. Wild,
Phys. Rev. C {\bf 82}, 024313 (2010).

\bibitem[{\citenamefont{Carlsson et~al.}(2008)\citenamefont{Carlsson,
  Dobaczewski, and Kortelainen}}]{Carlsson:2008gm}
\bibinfo{author}{\bibfnamefont{B.~G.} \bibnamefont{Carlsson}},
  \bibinfo{author}{\bibfnamefont{J.}~\bibnamefont{Dobaczewski}},
  \bibnamefont{and}
  \bibinfo{author}{\bibfnamefont{M.}~\bibnamefont{Kortelainen}},
  \bibinfo{journal}{Phys. Rev. C} \textbf{\bibinfo{volume}{78}},
  \bibinfo{pages}{044326} (\bibinfo{year}{2008}).

\bibitem[{\citenamefont{Stoitsov et~al.}(2010)\citenamefont{Stoitsov,
  Kortelainen, Bogner, Duguet, Furnstahl, Gebremariam, and
  Schunck}}]{Stoitsov:2010ha}
\bibinfo{author}{\bibfnamefont{M.}~\bibnamefont{Stoitsov}},
  \bibinfo{author}{\bibfnamefont{M.}~\bibnamefont{Kortelainen}},
  \bibinfo{author}{\bibfnamefont{S.~K.} \bibnamefont{Bogner}},
  \bibinfo{author}{\bibfnamefont{T.}~\bibnamefont{Duguet}},
  \bibinfo{author}{\bibfnamefont{R.~J.} \bibnamefont{Furnstahl}},
  \bibinfo{author}{\bibfnamefont{B.}~\bibnamefont{Gebremariam}},
  \bibnamefont{and} \bibinfo{author}{\bibfnamefont{N.}~\bibnamefont{Schunck}},
  \bibinfo{journal}{Phys. Rev. C} \textbf{\bibinfo{volume}{82}},
  \bibinfo{pages}{054307} (\bibinfo{year}{2010}).

\bibitem[{\citenamefont{sadoudi}(2011)}]{sadoudi11thesis}
\bibinfo{author}{\bibfnamefont{J.}~\bibnamefont{Sadoudi}}, T. Duguet, J. Meyer and M. Bender
arXiv:1310.0854 (2013).
\bibitem{casas}
M.~Casas, J.~Martorell, E.~Moya de Guerra, and J.~Treiner,
\bibinfo{journal}{Nucl. Phys. A} \textbf{\bibinfo{volume}{473}},
  \bibinfo{pages}{429} (\bibinfo{year}{1987}).

\bibitem[{\citenamefont{Carlsson et~al.}(2010)\citenamefont{Carlsson,
  Dobaczewski, Toivanen, and Vesel\'y}}]{Carlsson:2009mq}
\bibinfo{author}{\bibfnamefont{B.}~\bibnamefont{Carlsson}},
  \bibinfo{author}{\bibfnamefont{J.}~\bibnamefont{Dobaczewski}},
  \bibinfo{author}{\bibfnamefont{J.}~\bibnamefont{Toivanen}}, \bibnamefont{and}
  \bibinfo{author}{\bibfnamefont{P.}~\bibnamefont{Vesely}},
  \bibinfo{journal}{Comput. Phys. Commun.} \textbf{\bibinfo{volume}{181}},
  \bibinfo{pages}{1641} (\bibinfo{year}{2010}).

\bibitem[{\citenamefont{Bennaceur}()}]{lenteur}
\bibinfo{author}{\bibfnamefont{K.}~\bibnamefont{Bennaceur}},
  \eprint{unpublished}.

\bibitem{bonche05}
P. Bonche, H. Flocard, P.-H. Heenen,
Comp. Phys. Comm. {\bf 171}, 49 (2005).

\end{thebibliography}
\end{document}